**Propensity towards Ownership and Use of Automated Vehicles: Who Are the Adopters? Who Are the Non-adopters? Who Is Hesitant?**

**Tho V. Le, Ph.D. (Corresponding author)**
Assistant Professor
School of Engineering Technology
Purdue University
401 N. Grant St. West Lafayette, IN 47907
Email: thovanle@purdue.edu
ORCID: 0000-0003-0166-2412

**Giovanni Circella, Ph.D.**
Director, 3 Revolutions Future Mobility Program
Institute of Transportation Studies
University of California, Davis, and
BOF Tenure-track Professor of Mobility
Department of Geography
Ghent University
1715 Tilia Street, Davis, CA 95616
Email: gcircella@ucdavis.edu

**ABSTRACT**

The objective of this study is to investigate automated vehicle (AV) adoption perceptions, including ownership intentions and the willingness to use self-driving mobility services. In this paper, we use data from the 2018 California Transportation Survey, and use K-means, a clustering technique in data mining, to reveal patterns of potential AV owners (and non-owners) as well as AV users (and non-users) of self-driving services. The results reveal seven clusters, namely *Multitaskers/ environmentalists/ impaired drivers*, *Tech mavens/ travelers, Life in transition, Captive car-users, Public/ active transport users, Sub-urban Dwellers*, and *Car enthusiasts*. The first two clusters include adopters who are largely familiar with AVs, are tech savvy, and who make good use of time during their commute. The last cluster comprise of non-adopters who are car enthusiasts. On the other hand, people who are *Life in transition*, *Captive car-users*, *Public/ active transport users*, and *Sub-urban dwellers* show uncertain perceptions towards being AV adopters. They are either pursuing higher education, having a busy schedule, supporting for sustainable society via government policies, or have a stable life, respectively. Insights from this study help practitioners to build business models and strategic planning, addressing potential market segments of individuals that are willing to own an AV vs. those that are more inclined to use self-driving mobility services. The "gray" segments identify a latent untapped demand and a potential target for marketing, campaigns, and sales.

## INTRODUCTION

The improvement of technology and the ease to access the internet have fueled up revolutions in transportation, such as vehicle automation and connectivity. Consequently, significant changes in transportation are expected in the future. In particular, car ownership and the use of fleet-based mobility services are the two areas that will be significantly affected by the introduction of automated vehicles (AVs).

The level of automation is classified on a 5-level scale (*1*). Currently, no Level 4 or Level 5 models are available on the market. Some models of Tesla are considered the most advanced models in the automation market, i.e. the model 3, model S, model X, and model Y. Interestingly, while car sales have declined (in favor of larger SUVs and pick-up trucks), sales from the energy-efficient electric Tesla cars significantly increased in the recent years in the US, and more recently in other international markets. **Figure 1** shows Tesla sales and its market share in the US from 2014 to 2021. The most significant increase is observed in 2021. In fact, Tesla 3 is currently the best-selling vehicle among Tesla's three models (*2*).

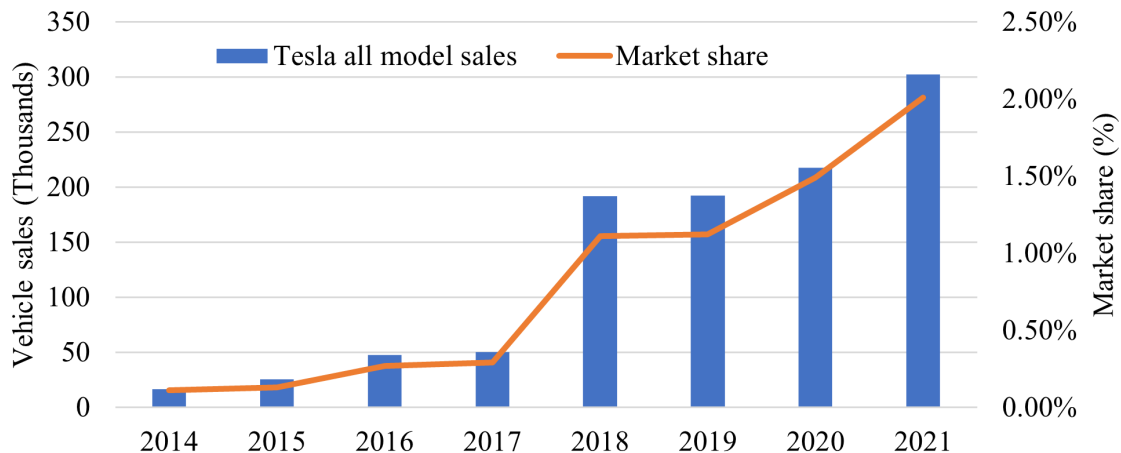

**Figure 1. Tesla sales and its market share in the US from 2014 to 2021 (*3*)**

Regarding the use of new mobility services, the easier access to Avs as well as the booming of emerging transportation services (e.g. ridehailing) are positioned to change the way people travel in cities through offering the benefits of real-time matching of demand and supply and additional mode choice options. Transportation network companies (TNCs), such as Uber and Lyft, have already teamed up with car manufacturers (e.g. BMW and Daimler) to prepare for the disruption of automation in the ridehailing service industry (*4*). One main goal of this collaboration is to provide ridehailing services using automated vehicle fleets. This business model will potentially change not only travel patterns and mode choice behaviors, but it is likely the bring heavy impacts on vehicle ownership preferences and intentions.

The transportation sector has already received heavy investments aimed at the development (and deployment) of new technologies and services in the last few years. The total global capital funding for travel and molility start-ups in 2014 and 2021 was $7.95 billion and $43.98 billion, respectively (*5*). Before Covid-19 pandemic, major capital was invested in some popular areas, such as TNCs, logistics, and freight transport firms (*6-9*). However, there are several areas lacking fundamental understanding, such as: (i) the potential markets for AVs and self-driving services; and (ii) the considerations that associate the success of AV adoptions in terms of ownership and use of self-driving services. As such, this study is expected to contribute to the following areas:

- Reveal segmentations of people who are hesitant to be AV owners and/or users of self-driving services. This group of people is on a fence, therefore, any small change in AV sale policy, self-driving services' strategy, or governmental policy is likely to give significant influence on decisions of the members of this group.
- Provide a better understanding about potential markets for AV sales and usage by answering questions such as: Who is likely to own an AV? Who is unlikely to be an AV owner? And who are the users and the non-users of self-driving mobility services?
- Offer knowledge of influencing factors, such as the technology usage, living lifestyle, work and life status, attitudinal factors, perceptions about AVs, and social demographic characteristics on the AV ownership and the use of self-driving services.

This paper is organized into seven sections. The first section introduces the topic and the need for this study. The second section summarizes previous studies and presents gaps from related literature. The third section describes the methodology used in this study. The fourth section presents the data source and its descriptive analysis. The fifth section shows results and provides insights. The sixth section discusses significant levels of influencing variables as well as the importance to unlock the latent AV demand. Finally, the paper concludes with the discussion of several avenues for future research in the final section.

## LITERATURE REVIEW

In this section, related studies in two main areas, namely AV ownership and self-driving services, are summarized. There are a few studies covering either one or both areas, with most of them only focusing on whether respondents are likely to own AVs or whether they are interested in using self-driving mobility services.

AV ownership has recently got increasing attention from researchers and practitioners. One study (*10*) summarizes the literature and categorizes influencing factors on AV ownership into several categories, namely socio-demographic characteristics and current behaviors. Sociodemographic variables include gender (*11, 12*), age (*12*), and income (*11, 12*). Whereas, the current behaviors includes vehicles miles traveled (*11, 12*)), car sharing (*12*), current vehicle autonomy level (*11*), and number of past crash experiences (*12*). More specifically, Kyriakidis et al. (*11*) use data collected from 109 countries and identifies factors with a positive effect on the likelihood of adopting an AV, including being male, having higher income, driving longer distances, and higher levels of automation. On the other hand, Bansal et al. (*12*) presents results from an online survey identifying positive determinants in being male, young and having higher income.

Furthermore, Nair et al. (*13*) employ order probit models to analyze the 2014 and 2015 data collected from the Puget Sound Regional Travel Study. This paper reveals that being male, driving alone on the commute, and household structure all affect AV ownership decisions. Whereas, the work from Lavieri et al. (*14*) using the same dataset reports that younger, higher educated people, and technologically-savvy groups are found to have higher AV ownership intentions.

There are several studies aiming to understand potential market segments of self-driving services. The factors that influence the interest in using self-driving mobility services include socio-demographic characteristics, individual attitudes, current behaviors, and trip attributes (*10*). Regarding socio-demographic characteristics, significant variables are gender (*15-17*), age (*18*), income, education level, and number of children in the household (*17*). Attitudinal variables include technology awareness (*18*), loss of control and sensation seeking (*16*), and data privacy concerns (*17*). Current behavior variables include current vehicle autonomy level (*17,18*). Trip attributes comprise of trip distance (*19*), and whether travel happens on highways or in congested traffic (*12*).

Both Shoettle et al. (*15*) and Payre et al. (*16*) indicate that males are more inclined to use self-driving services. Zmud et al. (*17*) report positive determinants including being male, having higher income, holding a graduate degree, and living in a household without children. Interestingly, the study of Nair et al. (*13*) show that people who are male, older, non-working, and without smartphones are more likely to use an AV as a taxi with a backup driver. Summarizing stated preference and choice studies, the work of Gkartzonikas et al. (*20*) presents nine factors affecting the intention to use AVs, including the "level of awareness," "consumer innovations," "safety," "trust of strangers," "environmental concerns," "relative advantage, compatibility, complexity," "subjective norms," "self-efficiency," and "driving related seeking scale."

Still, the research on market segments for AV ownership and use of self-driving services is limited. Therefore, this work aims to fill several gaps in the literature including:

- Only a few studies focus on both sides of the market: owners and non-owners of AVs as well as users and non-users of the self-driving services. The available literature also has not yet investigated people who are hesitant towards AV adoptions. We aim to uncover these market segments--the "gray" areas of AV adoption--which potentially are niche markets.
- Prior research on AV ownership mostly relies on statistical models, descriptive analysis, or discrete choice models. This approach is subjective in selecting explanatory variables that sometimes may lead to miss-identify and/or miss-report unexpected influencing factors.
- There is a limited number of studies that use data mining techniques for understanding AV ownership and interest in the use of self-driving services. Data mining techniques are useful to analyze a large dataset in order to obtain hidden knowledge and better insights.

The next section presents the research approach and methods we use to reveal market segments for AV ownership and the use of self-driving services.

## METHODS

Addressing the aforementioned gaps in literature, our overall approach uses a data mining technique to identify people who have neutral or mixed preferences towards the ownership and use of AVs. Moreover, this study clusters potential AV owners and non-owners as well as users and non-users of self-driving services. **Figure 2** presents a conceptual framework of this study.

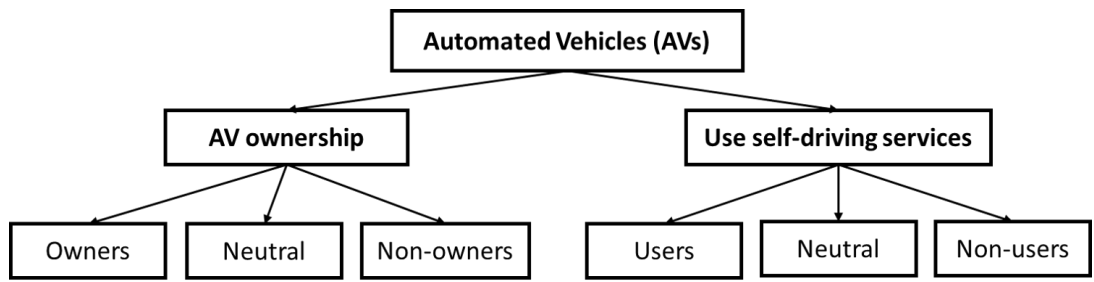

**Figure 2. Conceptual framework**

There are several common methods for clustering un-labeled data, such as mean shift and K-means. Mean-shift method is usually employed for a dataset that is unknown for the number of clusters in advance, and has uneven cluster sizes and a non-flat geometry (*21*). The K-means method is generally used for clustering a dataset that is known (or will be computed) for the number of clusters, and has an even cluster size and a flat geometry (*21*). In this study, K-means method is used due to its simplicity and powerfulness.

"The K-means algorithm clusters data by trying to separate samples in n groups of equal variances, minimizing a criterion known as the inertia or within-cluster sum-of-squares [...] This algorithm requires the number of clusters to be specified. It scales well to large number of samples and has been used across a large range of application areas in many different fields" (*21*).

K-means computation uses the following equations (*21*). Given an initial guess of $K$ clusters, the candidate cluster centroids $\mu_j, j = [1, |K|]$ are generated. Then, the algorithm repeats the two following steps until the process converges. Step one is to compute an objective value of the program of $n$ data points. The objective of K-means algorithm is to select centroids in order to minimize the within-cluster sum-of-squares (wcss) which is presented in **Equation** 1. As such, each data point $x_i$ is assigned to a cluster if it is closer to that cluster's centroid than the other centroids.

$$\sum_{i=1}^{n} \min_{j}\left(\left\|x_i - \mu_j\right\|\right)^2 \qquad (1)$$

The second step is to update the centroid $\mu_j$ which is determined by **Equation** 2.

$$\mu_j = \frac{1}{|S_j|} \sum_{x_j \in S_j} x_j \qquad (2)$$

where $S_j$ is a set of data points for each $j^{th}$ cluster.

To find the optimal number of clusters, we use the elbow method. Based on the values of the wcss computed for each number of clusters, we plot the wcss against its corresponding number of clusters. From the plot, it is usually easy to identify the optimal number of clusters, which corresponds to the *elbow* point in the plot (*22*). Otherwise, we can compute distances from points to a based line that connects the first and the last points of the curve (*22, 23*). K-means is a distance-based method and the natural presentation of variables in our dataset is in different scales. Therefore, all variables were normalized and included in the analysis with the assumption that they play a similar role in the dataset (e.g. they are "weighted" in a similar way in the definition of the optimal cluster solution).

## DATA AND DESCRIPTIVE ANALYSIS

This research utilizes the 2018 California Transportation Survey data. The data were collected in California from June to November 2018 using both online- and paper-based surveys. After cleaning, the number of total valid cases is 2,336. For more information on the survey design and data collection for this project, see the project report from Circella et al. (*24*).

In the survey, respondents were requested to provide information on their preferences, lifestyles, and adoption of new transportation technology and services, among other groups of variables. Further, in the *future mobility* section of the survey, respondents were asked about their future expectations on car ownership, interesting in purchasing alternative-fuels vehicles, use of car-sharing and other shared mobility options, and the familiarity with the AV/self-driving concept. After that, respondents were provided more information including definitions of self-driving vehicles, its features, and pictures for visualization. Respondents were then asked to rate on a scale from "strongly disagree" to "strongly agree" multiple statements regarding safety and travelling enjoyment in an era of self-driving vehicles. Respondents were also asked to indicate their likelihood to use a self-driving vehicle for various activities/purposes. Summary statistics for some main variables are presented in **Table 1**

**Table 1. A summary of respondents' socio-demographic characteristics**

| Attributes | Mean/ Standard Deviation or Percentage |
|---|---|
| Age | 44.76/ 14.69 |
| Gender: Male; Female; Transgender | 48.50%; 51.30%; 0.20% |
| Race: Black; Native American; Asian; White; Multiple races; Others | 3.60%; 2.30%; 14.30%; 74.30%; 4.20%; 1.40% |
| Educational background: Some grade/ high school; Completed high school or GED; Some college/ technical school; Bachelor' degree; Graduate degree; Professional degree | 1.30%; 7.40%; 30.00%; 37.80%; 18.70%; 5.00% |
| Annual household income: Less than $25,000; $25,000 to $49,999; $50,000 to $74,999; $75,000 to $99,999; $100,000 to $149,000; $150,000 or more | 8.00%; 17.40%; 17.10%; 15.40%; 20.80%; 21.2% |

This study is built on the responses to questions assuming the availability of self-driving vehicles. Respondents were asked how likely they would be to own a personal self-driving vehicle and/or use self-driving services (such as driverless taxi). Respondents were asked to rate their interest in the following:

- "Be one of the first people to buy a self-driving vehicle."
- "Eventually buy a self-driving vehicle, but only after these vehicles are commonly used."
- "Use a driverless taxi alone or with others I know."
- "Use driverless taxi or shuttle with other passengers who are strangers to me."

In this study, we employ variables that can be grouped in three main categories, as shown in Table 2. The attitudinal factors were obtained from a factor analysis of respondent's opinions on various topics. **Table 2** reports the variable names, their meaning, and their normalized mean and standard deviation.

**Table 2. Variables, its meaning, and normalized mean and standard deviation**

| Variables | Meanings | Normalized mean/standard deviation |
|---|---|---|
| **Socio-demographic characteristics** | | |
| Age | Actual age | 0.46/ 0.15 |
| Gender | Gender | 0.51/ 0.17 |
| Education | Highest educational background | 0.63/ 0.17 |
| Household_Income | Household income | 0.65/ 0.27 |
| Vehicle_Ownership | Vehicle ownership status | 0.95/ 0.17 |
| **Household size** | | |
| HouseholdSize_Total | Total number of people live in the household | 0.23/0.12 |
| **Residence type and housing tenure** | | |
| NeighborhoodType_Grewup | Characterize the area where grew up | 0.73/0.23 |
| NeighborhoodType_Current | Characterize the current living area | 0.78/0.22 |
| NeighborhoodType_Future | Characterize the area to live in the future | 0.72/0.24 |
| Housing_Tenure | Rent, own, or provided housing by someone | 0.75/0.31 |
| **Key life-events** | | |
| Move_NewPlace | Move to a different address | 0.25/0.31 |
| Change_JobLocation/NewJob | Start new job/change job (or job location) | 0.24/0.29 |
| Retire/LeaveJob | Retire/leave job | 0.13/0.22 |
| Start_Studies | Start studies | 0.09/0.19 |
| Graduate/End_Studies | Graduate/finish/end studies | 0.10/0.20 |
| Married/SeriousRelationship | Get married/enter a serious relationship | 0.12/0.21 |
| Separated/Divorced | Get separated/divorced/widowed | 0.02/0.09 |
| Have/Adop_aChild | Have/adopt a child | 0.09/0.19 |
| HouseholdMember_MoveOut | Have some members of my household (*e.g.* children) move out | 0.11/0.22 |
| Someone_MoveIn | Have somebody (*e.g.* family member or partner) move in with | 0.08/0.18 |
| **Current travel choices for commute trips** | | |
| CommuteFreq_CarAlone | Car, alone | 0.80/0.31 |
| CommuteFreq_CarOthers | Car, with others (*e.g.* carpooling) | 0.26/0.33 |
| CommuteFreq_Shuttle | Work/school-provided bus or shuttle | 0.08/0.18 |

| Variables | Meanings | Normalized mean/standard deviation |
|---|---|---|
| CommuteFreq_Bus | Public bus | 0.18/0.22 |
| CommuteFreq_Lrail/Subway | Light rail/tram/subway (*e.g.* BART, LA Metro) | 0.13/0.23 |
| CommuteFreq_Train | Commuter train (*e.g.* Amtrak, Caltrain) | 0.08/0.15 |
| CommuteFreq_Taxi | Taxi (*e.g.* Yellow Cab) | 0.14/0.12 |
| CommuteFreq_Ridehailing | Ridehailing (*e.g.* Uber, Lyft) | 0.18/0.17 |
| CommuteFreq_SharedRidehailing | Shared ridehailing (*e.g.* UberPOOL, Lyft Line) | 0.14/0.16 |
| CommuteFreq_Bike | Bicycle or e-bike | 0.14/0.18 |
| CommuteFreq_Walk | Walk or skateboard | 0.20/0.26 |
| CommuteFreq_Other | Other mode | 0.08/0.13 |
| **Current travel choices for leisure/shopping/social trips** | | |
| LeisureFreq_CarAlone | Car, alone | 0.62/0.29 |
| LeisureFreq_CarOthers | Car, with others (*e.g.* carpooling) | 0.41/0.32 |
| LeisureFreq_Shuttle | Carsharing (*e.g.* Zipcar, Car2Go) | 0.05/0.14 |
| LeisureFreq_Bus | Public bus | 0.07/0.17 |
| LeisureFreq_Lrail/Subway | Light rail/tram/subway (*e.g.* BART, LA Metro) | 0.08/0.17 |
| LeisureFreq_Train | Commuter train (*e.g.* Amtrak, Caltrain, Metrolink) | 0.04/0.12 |
| LeisureFreq_Taxi | Taxi (*e.g.* Yellow Cab) | 0.04/0.13 |
| LeisureFreq_Ridehailing | Ridehailing (*e.g.* Uber, Lyft) | 0.12/0.19 |
| LeisureFreq_SharedRidhailing | Shared ridehailing (*e.g.* UberPOOL, Lyft Line) | 0.07/0.16 |
| LeisureFreq_Bike | Bicycle or e-bike | 0.09/0.20 |
| LeisureFreq_Walk | Walk or skateboard | 0.20/0.28 |
| LeisureFreq_Other | Other (please specify): | 0.08/0.13 |
| **Current travel characteristics** | | |
| Car_VMT | Miles driven per week | 0.07/0.08 |
| Limit_to_Drive | physical or other personal conditions that prevent or limit from driving | 0.04/0.18 |
| **Attitude towards AVs** | | |
| AV_Familiarity | Awareness of or familiarity with the concept of AVs/self-driving vehicles. | 0.70/ 0.20 |
| AV_Travel_Enjoyment | A self-driving vehicle would enable me to enjoy traveling more (e.g. watching the scenery). | 0.67/ 0.28 |
| AV_Misscontrolling_Car | I would miss the joy of driving and of being in control. | 0.70/ 0.26 |
| AV_Safety_Anxiety | I would be concerned about the safety of the occupants of the self-driving vehicle. | 0.83/ 0.22 |
| AV_Drives_at_Slow_Speed | I would accept longer travel times, so the self-driving vehicle drives at a speed low | 0.64/ 0.26 |

| Variables | Meanings | Normalized mean/standard deviation |
|---|---|---|
|  | enough to prevent unsafe situations for pedestrians and bicyclists. |  |
| No_need_AV | I do not see any need for self-driving vehicles. | 0.58/ 0.27 |
| AV_Travel_When_Tired | More often travel even when I am tired, sleepy, or under the influence of alcohol/medications. | 0.63/ 0.29 |
| Work_on_AV_Less_Office_hr | Reduce my time at the regular workplace and work more in the self-driving car. | 0.49/ 0.27 |
| **Attitudinal variables** | | |
| Sustainable_via_Gov_Policy | Government should raise the price of gasoline to reduce the negative impacts on the environment, and to provide funding for better public transportation. The government should put restrictions on car travel in order to reduce congestion. | 0.35/ 0.25 |
| Tech_Maven | I like to be among the first people to have the latest technology. I like trying things that are new and different. I would/do enjoy having a lot of luxury things. | 0.54/ 0.17 |
| Car_Enthusiast | I definitely want to own a car. | 0.72/ 0.18 |
| Suburbia_Good_Living_Place | I prefer to live in a spacious home, even if it is farther from public transportation and many places I go. | 0.55/ 0.22 |
| Busy_Car_Dependent | My schedule makes it hard or impossible for me to use public transportation. | 0.64/ 0.22 |
| Commute_Multitaskers | My commute is a useful transition between home and work (or school). I try to make good use of the time I spend commuting. | 0.51/ 0.15 |
| Eco_Minimalism_Lifestyle | I am committed to an environmentally-friendly lifestyle. | 0.56/ 0.17 |
| Life/Career_Adriftingness | I'm still trying to figure out my career (e.g. what I want to do, where I'll end up). | 0.41/ 0.17 |
| Car_Utilitarian | The functionality of a car is more important to me than its brand. | 0.65/ 0.20 |
| **Opinion about the ownership and usage of AVs** | | |
| AV_Own_First_to_Buy | Be one of the first people to buy a self-driving vehicle. | 0.40/ 0.25 |
| AV_Own_Follow_to_Buy | Eventually buy a self-driving vehicle, but only after these vehicles are commonly used. | 0.60/ 0.28 |
| AV_Use_Taxi | Use a driverless taxi alone or with others I know. | 0.53/ 0.28 |
| AV_Use_Shared_Taxi | Use a driverless taxi or shuttle with other passengers who are strangers to me (like UberPOOL). | 0.46/ 0.26 |

The respondents were also asked about their awareness of or familarity with the concept of self-driving vehicles. The results are presented in Figure 3. About 34% of respondents have never hear of it or not familiter with it. This unfamiliarity with the concept may influent on their decisions in AV adoptions. On the other hand, there are about 47% of respondents familiar with the concept while approximately 19% of people said they have great understanding of the self-driving concept.

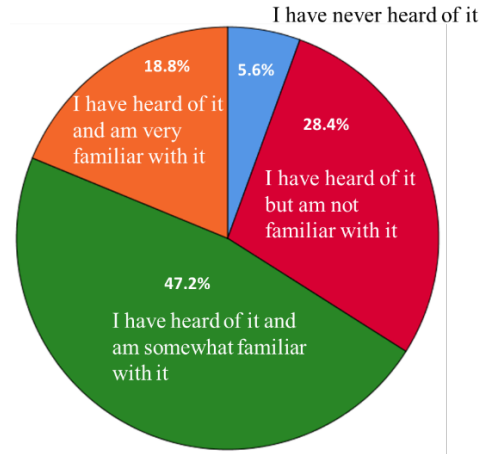

**Figure 3. The share of respondents' awareness of or familiarity with the concept of self-driving vehicles**

The survey also collects the type of fuel that respondents currently use for their car. Gasoline stands out the most popular fuel use in over 87% of respondents' cars. At the second place, hybrid cars comprise about 6.5%. Plug-in hybrid, battery electric, diesel, flex-fuel vehicles, and other fuels are less commonly used. The results are illustrated in **Figure 4**. Moreover, the survey also gathers information on how respondents used their time during their recent commute trip as presented in **Figure 5**. Most of respondents reported they either used their smartphone or talked on their phone during their last trip. Interestingly, about 18% of respondents communicated with other travelers. Some people reported they worked on a laptop or tablet or used non-electronic items (e.g., read a book or newspaper, or played cards). This information of time use during trips supports AV design of seating and interior devices that aim to maximize users' experiences while using AVs.

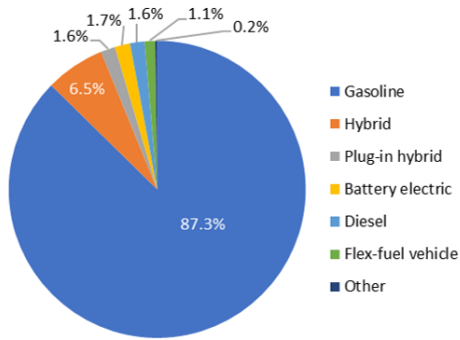

**Figure 4.** Distribution of current fuel-type usages.

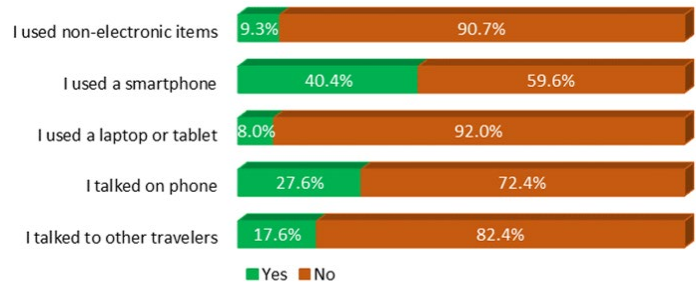

Figure 5. Time spend during commute trips.

Furthermore, it is interesting to explore how the future transportation technologies and services effect on people lifestyle. As such, we asked people's perceptions of when AVs will fully automate, and if AVs were available, where do people live. Over 51% of people said the fully AVs will be available in less than 20 years whereas about 35% of people thought it will take 21 years or more. Noticeably, around 14% of respondents did not think it is feasible to have all cars automated in the future. The results are illustrated in Figure 6. Moreover, people also provided their perceptions on where they live in an era of AVs. Surprisingly, around 75% of respondents said they will not move to a new place even if AVs available. There are over 15% and 10% of respondents said they will move closer and farther, respectively, to/from the location that they travel to most often. Given that there is only about 19% of respondents are familiar with AV concept so some benefits of owning AVs or using AV services have not yet cleared to majority of respondents. It is also possible that there are some reasons associated to respondents' perception of staying at their current place and should be investigated in the future study. For example, people may own a great house that they dream for; or people may want to live close to their family members and relatives; or they are living in an excellent community/area that they do not want to leave.

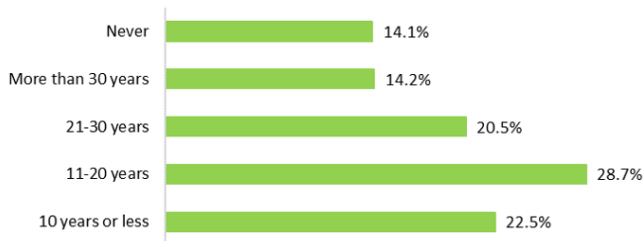

**Figure 6. Expected time to have fully AVs.**

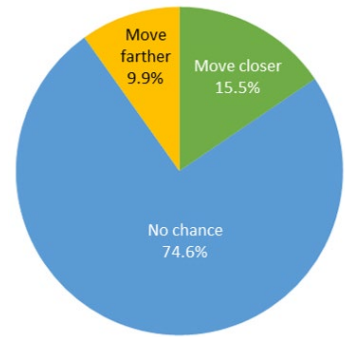

**Figure 7. Changes in living location to the most travel places when AVs were available.**

## RESULTS AND INSIGHTS

This section presents results and insights from the analysis of respondents' perceptions on how likely they are to be AV adopters. We organize the discussion around three levels of perceptions: likely, hesitant, and unlikely (to adopt AVs and/or self-driving mobility services). Seven clusters are identified: *Multitaskers/ Environmentalists/ Impaired drivers*, *Tech mavens/ travelers, Life in transition, Captive car-users, Public/ active transport users, Sub-urban dwellers*, and *Car enthusiasts*. **Figure 8** presents ownership and usage perceptions together with its corresponding determinants under seven clusters.

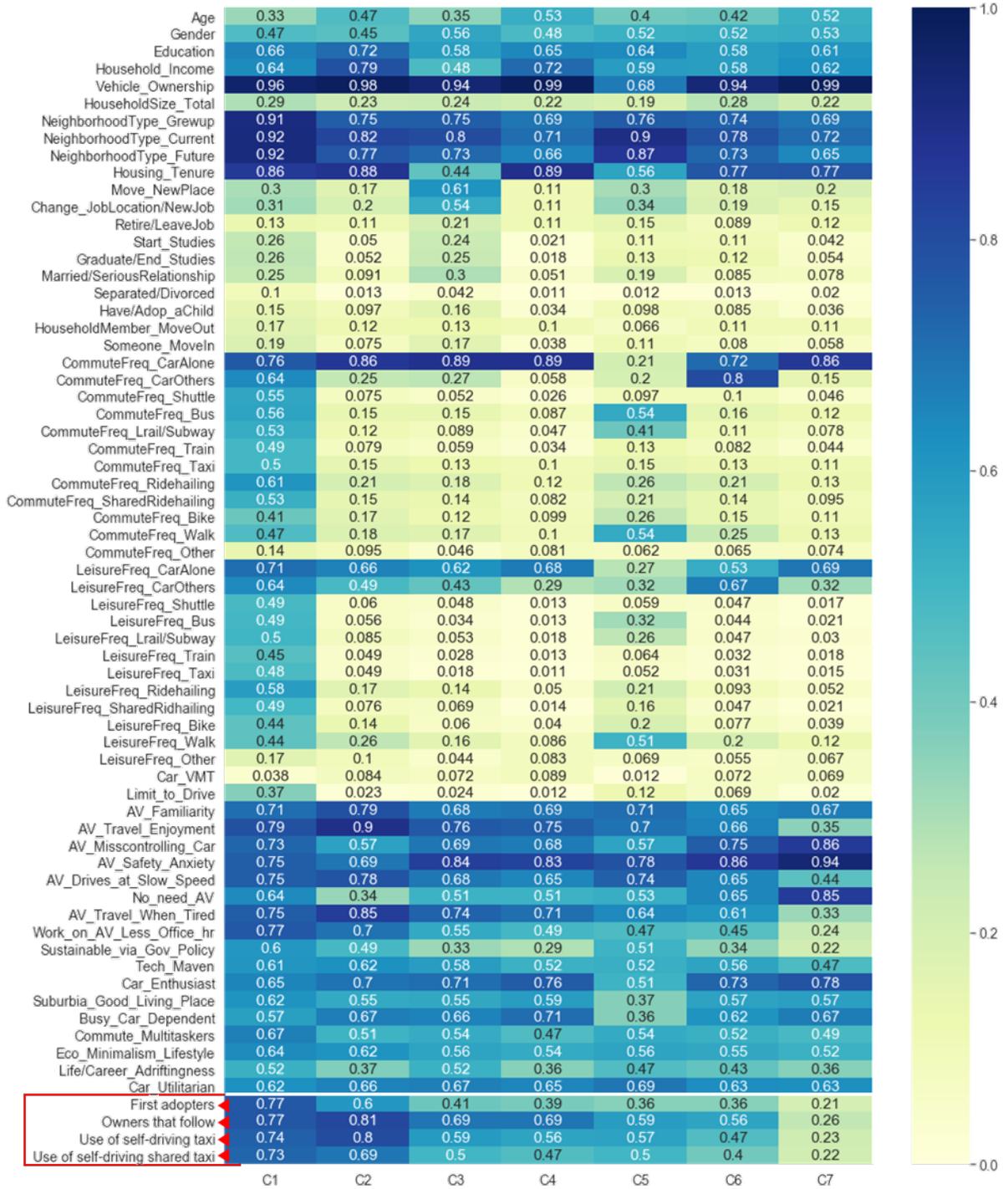

**Figure 8. Seven clusters showing ownership and usage perceptions with its determinants. C1:** *Multitaskers/ environmentalists/ impaired drivers*; **C2:** *Tech mavens/ travelers*; **C3:** *Life in transition*; **C4:** *Captive car-users*; **C5:** *Public/ active transport users*; **C6:** *Sub-urban dwellers*; **and C7:** *Car enthusiasts*. **The X axis reports the cluster names, and the Y axis shows the variable names. The reported values are the center-values for each variable (by cluster) (with darker colors identifying higher values, and lighter colors identifying lower values). All variables are normalized and included in the interval [0,1]. Unlikely adopter-group has values within [0,0.5]; neutral group has**

**values within (0.5,0.7); Likely adopter-group has values within [0.7,1] for the four adoption variables reported at the bottom of the figure.**

**Who is likely to be an AV adopter?**

Not surprising, likely AV adopters are very familiar with the concept of AVs/ self-driving vehicles. Adopters are men who earn advanced education, and generally want to work on AVs so they can shorten working time in their office thanks to AVs. Interestingly, *multitaskers/ environmentalists/ impaired drivers* include early adopters who are young men, want to make good use of the time during their commute, have minimalism lifestyle, and have physical or personal conditions that prevent them from driving. These people frequently use public transport, taxi, ridehailing, bike, and believe that the government should introduce transportation-related policies that are aimed to regulate transportation and increase sustainability. They grew up and currently live in urban areas. Their average household size is around 3.5 which is the largest among seven clusters. *Tech mavens/ travelers*, however, include individuals who are more likely to own AVs after these vehicles are commonly used. These people are also interested in using self-driving services, even if they are more hesitant to be early owners. Those individuals are middle-age men, tech mavens, and have very high income. The members of this cluster enjoy travelling by using AVs, and expect benefits from the use of AVs as they will allow them to travel even when they are tired or under the influence of alcohol/medications. The *multitaskers/ environmentalists/ impaired drivers* and *tech mavens/ travelers* include approximately 21% of the 2,336 cases in the sample. Our findings on adopters align pretty well with previous studies of (*12, 14, 15*) with regard to age, gender, tech savviness, education, and income. In addition, we identify new relevant variables that affect the likelihood of adopting AVs are household size, residence type, current travel choices, and current travel characteristics. We also reveal driving impaired people who are a niche market for AVs.

**Who is more hesitant?**

The AV usage is not yet common as well as knowledge about AVs is more limited for individuals in the intermediate clusters. That lower awareness is likely to influence these individuals' decisions. In fact, our data show that about 56% of respondents in this group are still hesitant about owning an AV or using self-driving services. The four clusters in this intermediate group are those that are *transition in their life*, *captive car-users*, the *public/active transport users*, and *sub-urban dwellers*.

Those who are *transition in their life* have little interest in being an early AV owner and a shared user of self-driving services. However, they are reluctant of being an AV owner at a later time and in using self-driving taxi without other users. This group includes young low-income and low education individuals who have limited knowledge about self-driving vehicles. Some of them have got out of a relationship. They just graduated and moved to a new place in suburban area to pursue their higher education. They are likely rent a place to live.

*Captive car-users* show very mixed perceptions: they are rather unlikely to be early AV owners, hesitant to be an owner at a later time, and a user of self-driving services without other users. They are predominantly senior people who grew up and are living in suburban or small cities. They have a busy schedule that makes it hard or impossible for them to use public transportation. Interestingly, on average, they have high income and own their house.

*Public/active transport users* are not a potential early AV owner. They are hesitant to be an AV owner even after AVs are commonly used and to be a user of self-driving mobility services. In most cases, these individuals are in their middle age, grew up and currently live in urban areas. They like to use public transport, and they walk frequently, possibly to access and egress to/from public transport facilities. These middle-age people are not open for longer commutes in order to live in a more spacious home.

*Suburban dwellers* are not interested in being the first AV owners and hesitant to be followers to buy AVs. They are also hesitant to use self-driving taxi but not the shared-taxi services. This cluster includes middle age people who frequently travel with their children, possibly because of sending and picking up their children from school and activities. These families currently live in suburban areas and have an average of 3.3 people in their household. They tend to have a stable job and ordinary life.

These four clusters comprise people who have mixed perceptions or are rather uninterested in being AV adopters. Our findings identify a segment of the population that can be potential future markets for both AV sales and TNC services.

**Who is unlikely to be an AV adopter? To what extent can we change their propensities?**

There is one distinct cluster including unlikely adopters: *car enthusiasts*. People in this group are non-tech mavens and worry about the occupants' safety when using AVs. Those people account for approximately one fourth of the total sample.

Those individuals show a limited understanding about self-driving vehicles/ AVs. They grew up and currently live in suburban areas or small towns. *Car enthusiasts* include senior people who love driving and controlling a car. They frequently drive alone for commute, leisure, shopping, and other purposes. The findings on the segments of non-adopters are interesting and distinctive, and identify groups that will be among the last to adopt self-driving technology.

**CLUSTERING PERFORMANCE EVALUATION**

This section presents performance evaluation approaches that validates seven optimal clusters. In fact, there is no solid evaluation matrix to evaluate the outcome of unsupervised clustering. In this study, we employ elbow method to assess the model performances. Figure 9 illustrates a plot of wcss again corresponding cluster. The optimal cluster is seven which its wcss has the longest distance (i.e. *d*=6.62) to the line connecting wcss of two clusters and 20 clusters.

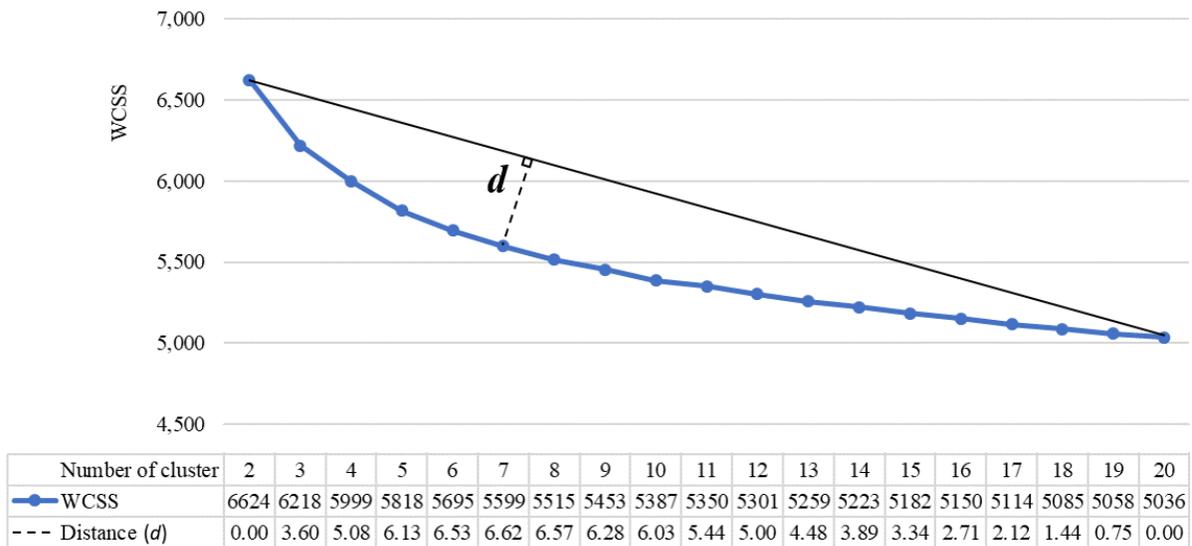

| Number of cluster | 2 | 3 | 4 | 5 | 6 | 7 | 8 | 9 | 10 | 11 | 12 | 13 | 14 | 15 | 16 | 17 | 18 | 19 | 20 |
|---|---|---|---|---|---|---|---|---|---|---|---|---|---|---|---|---|---|---|---|
| WCSS | 6624 | 6218 | 5999 | 5818 | 5695 | 5599 | 5515 | 5453 | 5387 | 5350 | 5301 | 5259 | 5223 | 5182 | 5150 | 5114 | 5085 | 5058 | 5036 |
| Distance (*d*) | 0.00 | 3.60 | 5.08 | 6.13 | 6.53 | 6.62 | 6.57 | 6.28 | 6.03 | 5.44 | 5.00 | 4.48 | 3.89 | 3.34 | 2.71 | 2.12 | 1.44 | 0.75 | 0.00 |

**Figure 9. Optimal cluster identification**

**DISCUSSIONS**

**Factors associated with propensities towards AV adoption and use of self-driving services**

This section discusses how various factors are associated with respondents' propensities towards ownership and usage of AVs. There are three types of variables in our study: variables that data collected from the use of the Likert scale of 5 (i.e. Attitude towards AVs, and Attitudinal variables); continuous variables (i.e. Age, HouseholdSize_Total, and Car_VMT); and categorical variables (i.e. other variables of Socio-demographic characteristics, Residence type and housing tenure, Key life-events, Current travel choices for commute trips, Current travel choices for leisure/ shopping/ social trips, and Limit_to_Drive). The association of those variables is decided by comparing center values of the variable in the cluster to that of the same variable in the other clusters, in corresponding to the overall scale of that variable. In general, the following notations and corresponding meanings are used in Table 3: (+): positive association (center values equal or larger than (mean + standard deviation/4)); (-): negative association (center values equal or less than (mean- standard deviation/4); (0): neutral association (center values between (mean - standard deviation/4) and (mean + standard deviation/4)); (+/-): unclear association (center values get different associations). The variables' mean and standard deviation are presented in Table 2.

**Table 3** summarizes the main direction of the association between several variables in the sample and the propensity to adopt and use AVs. In general, there are two groups of similar characteristics which are a group of early owners and shared taxi users, and a group of later owners and taxi users. There are many common variables that are positively associated with the ownership and usage of AVs. These include education, income, vehicle ownership, familiarity with AV concept, enjoying traveling more with AVs, traveling more often, using time productively on AVs, supporting sustainable travel, tech savviness, and environmental-friendly lifestyle. Adopters are likely to be men. On the contrary, younger people are more likely to be early adopters and use shared AV taxi, while age shows an unclear relationship with AV-related intentions of later owners and taxi users. The variables of key life-events and current travel choices are either neutral associations or positive for early adopters and shared-taxi users, whereas they are unclear for the remaining adopters. The factor's signs reported in **Table 3** are the results of two clusters of adopters, i.e. *multitaskers/ environmentalists/ impaired drivers* and *tech mavens/ travelers*.

Regarding AV ownership, younger people are more likely to be early adopters, whereas both youngers and seniors are more likely be AV owner only after the AV technology is well established in the market. In addition, advanced-degree holders, higher income people are more open to be owners of AVs, while the women do not seem to have direct association with being AV owners.

With respect to the use of self-driving services, young people are more likely to use shared self-driving mobility services, whereas people at different age are willing to use self-driving taxi services. This probably is due to younger people's openness to the new and innovative technology. Moreover, the education level and household income show a positive impact on the use of taxi services.

**Table 3. Factors associated with the propensities towards ownership and use of AVs**

| Factor | AV ownership | | Use of self-driving services | |
|---|---|---|---|---|
| | Early owners | Later owners | Taxi | Shared taxi |
| Age | (-) | (+/-) | (+/-) | (-) |
| Gender | (-) | (-) | (-) | (-) |
| Education | (+) | (+) | (+) | (+) |
| Household_income | (+) | (+) | (+) | (+) |
| Vehicle_ownership | (+) | (+) | (+) | (+) |
| HouseholdSize_Total | (+) | (+/-) | (+) | (+) |
| NeighborhoodType_Grewup | (+) | (+/-) | (+) | (+) |
| NeighborhoodType_Current | (+) | (+/-) | (+) | (+) |
| NeighborhoodType_Future | (+) | (+/-) | (+) | (+) |
| Housing_Tenure | (+) | (+/-) | (+) | (+) |
| Move_NewPlace | (0) | (0) | (0) | (0) |
| Change_JobLocation/NewJob | (0) | (0) | (0) | (0) |
| Retire/LeaveJob | (0) | (0) | (0) | (0) |
| Start_Studies | (+) | (+/-) | (+/-) | (+) |
| Graduate/End_Studies | (+) | (+/-) | (+/-) | (+) |
| Married/SeriousRelationship | (+) | (+/-) | (+/-) | (+) |
| Separated/Divorced | (+) | (+/-) | (+/-) | (+) |
| Have/Adop_aChild | (+) | (+/-) | (+/-) | (+) |
| HouseholdMember_MoveOut | (0) | (0) | (0) | (0) |
| Someone_MoveIn | (+) | (0) | (0) | (+) |
| CommuteFreq_CarAlone | (+) | (+/-) | (+/-) | (+) |
| CommuteFreq_CarOthers | (+) | (+/-) | (+/-) | (+) |
| CommuteFreq_Shuttle | (+) | (+/-) | (+/-) | (+) |
| CommuteFreq_Bus | (+) | (+/-) | (+/-) | (+) |
| CommuteFreq_Lrail/Subway | (+) | (+/-) | (+/-) | (+) |
| CommuteFreq_Train | (+) | (+/-) | (+/-) | (+) |
| CommuteFreq_Taxi | (+) | (+/-) | (+/-) | (+) |
| CommuteFreq_Ridehailing | (+) | (+/-) | (+/-) | (+) |
| CommuteFreq_SharedRidehailing | (+) | (+/-) | (+/-) | (+) |
| CommuteFreq_Bike | (+) | (+/-) | (+/-) | (+) |
| CommuteFreq_Walk | (+) | (+/-) | (+/-) | (+) |
| CommuteFreq_Other | (+) | (+/-) | (+/-) | (+) |
| LeisureFreq_CarAlone | (+) | (+/-) | (+/-) | (+) |
| LeisureFreq_CarOthers | (+) | (+/-) | (+/-) | (+) |
| LeisureFreq_Shuttle | (+) | (+/-) | (+/-) | (+) |
| LeisureFreq_Bus | (+) | (+/-) | (+/-) | (+) |
| LeisureFreq_Lrail/Subway | (+) | (+/-) | (+/-) | (+) |
| LeisureFreq_Train | (+) | (+/-) | (+/-) | (+) |
| LeisureFreq_Taxi | (+) | (+/-) | (+/-) | (+) |
| LeisureFreq_Ridehailing | (+) | (+/-) | (+/-) | (+) |
| LeisureFreq_SharedRidhailing | (+) | (+/-) | (+/-) | (+) |
| LeisureFreq_Bike | (+) | (+/-) | (+/-) | (+) |
| LeisureFreq_Walk | (+) | (+/-) | (+/-) | (+) |
| LeisureFreq_Other | (+) | (+/-) | (+/-) | (+) |
| Car_VMT | (-) | (-) | (-) | (-) |
| Limit_to_Drive | (+) | (+/-) | (+/-) | (+) |
| AV_Familiarity | (+) | (+) | (+) | (+) |
| AV_Travel_Enjoyment | (+) | (+/-) | (+/-) | (+) |
| AV_Misscontrolling_Car | (+) | (+/-) | (+/-) | (+) |
| AV_Safety_Anxiety | (+) | (+/-) | (+/-) | (+) |
| AV_Drives_at_Slow_Speed | (+) | (+/-) | (+/-) | (+) |
| No_need_AV | (0) | (0) | (0) | (0) |
| AV_Travel_When_Tired | (+) | (+/-) | (+/-) | (+) |
| Work_on_AV_Less_Office_hr | (+) | (+/-) | (+/-) | (+) |
| Sustainable_via_Gov_Policy | (+) | (+/-) | (+/-) | (+) |
| Tech_Maven | (+) | (+) | (+) | (+) |
| Car_Enthusiast | (+) | (+/-) | (+/-) | (+) |
| Suburbia_Good_Living_Place | (+) | (+/-) | (+/-) | (+) |
| Busy_Car_Dependent | (0) | (0) | (0) | (0) |
| Commute_Multitaskers | (+) | (+/-) | (+/-) | (+) |
| Eco_Minimalism_Lifestyle | (+) | (+/-) | (+/-) | (+) |
| Life/Career_Adriftingness | (+) | (+/-) | (+/-) | (+) |
| Car_Utilitarian | (+) | (+/-) | (+/-) | (+) |

Legend: ▇ : (-)   ▇ : (+)   ▇ : (0)   ▇ : (+/-)

**Unlocking potential AV markets**

**FIGURE 10** summarizes the distribution of individuals in the sample, using the seven clusters identifies in the previous section. The cluster sizes suggest there are fewer adopters than non-adopters as visualized in an egg shape. Not surprising, a large proportion of respondents is between these two groups.

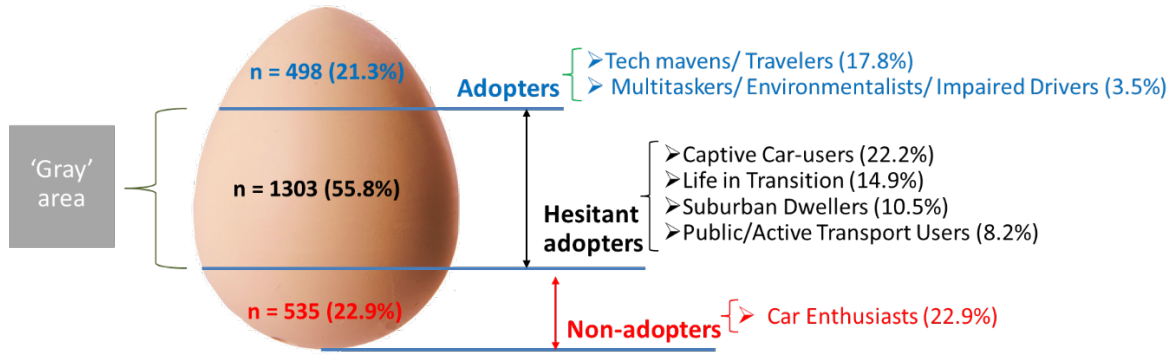

**FIGURE 10. Market segments and 'gray' areas. An egg shape identifies the smaller number of early adopters (i.e., 21.3%) than non-adopters (i.e., 22.9%) as well as larger proportion of hesitant adopters (i.e., 'gray' areas—55.8%).**

The segments of *hesitant adopters* and the people who have mixed perceptions are 'gray' areas: the individuals in these groups are on the edge of the market, and then be the potential target of campaigns and policies to their interest in adopting AVs. It is probably less costly to turn them to be customers than to convince non-adopters to change their mind. Product and service providers should investigate the needs of the individuals in the 'gray' areas and evaluate what features of the AV products and/or services these customers would prefer. Satisfying their needs is a key step to convert this latent demand of people having *life in transition, captive car-users, public/active transport users,* and *suburban dwellers* to potential customers.

Non-adopters are a segment that has the farthest distance to the market. They are car enthusiasts. Car enthusiasts can be a hard market to be tapped because their needs can difficulty be satisfied by AVs. Those people may not be willing to pay more for self-driving technology that they are not interested in using. They do not want to lose the excitement of being in control of their car. Nevertheless, in a short term, those people are a large market for selling manual-driving cars.

In conclusion, our study identifies the large latent demand waiting to be tapped in the AV market. The insights gained from this study can accelerate and guide the efforts to unleash non-customers and convert latent untapped demand to be customers. By studying their needs and involving them in creating products and services, a portion of the hesitant adopters in the "gray" area can turn into potential AV adopters.

**Characteristics of the three segments**

This section generalizes the characteristics of adopters, hesitant adopters, and non-adopters. Those characteristics are significant different across three segments. While the average age of early adopters and hesitant adopters is around 42-43 years old, the non-adopters are found to be about 8 years older. The average household income of early adopters is about 1.3 times higher than the other segments. There are about 85% of people own a car. Interestingly, early adopters enjoy driving and they drive more than the other people. The gasoline car-users are approximately 80%, 85%, and 91% for three segments, respectively. Surprisingly, around 62% of early adopters reported the increase or no change in the number of cars in their household in the last three years. This number is remarkably higher than those of the other two remaining segments. On the other hand, about 43%-47% of people said the number of cars in their household will not reduce in the next three years. Furthermore, people also reported about their preferences for buying or leasing an electric vehicle (EV). As expected, far more early adopters (i.e. 65.5%) are willing to buy or lease an EV in the next three years, whereas that of hesitant adopters and non-adopters are about 48% and 25%, respectively. Meanwhile, the propensities of using carsharing within the next year are in the similar trend which is about 37%, 10%, and 5% for early adopters, hesitant adopters, and non-adopters, respectively.

**Table 4. Adopters' characteristics**

|  | Adopters (n=498) | Hesitant Adopters (n=1,303) | Non-adopters (n=535) |
|---|---|---|---|
| Average age | 43.24 | 42.91 | 50.70 |
| Average annual household income ($) | $118,162 | $88,325 | $89,338 |
| Average travel distance (miles/week) | 153.00 | 140.22 | 137.82 |
| Car owners (%) | 93.40% | 85.10% | 95.50 |
| Gasoline car-users (%) | 79.50% | 85.30% | 91.00% |
| Increase/ keep the same total number of household cars in the last three years (%) | 62.10% | 52.80% | 48.80% |
| Increase/ keep the same total number of household cars in the next three years (%) | 43.60% | 47.00% | 45.60% |
| Willingness to buy or leasing EVs (%) | 65.50% | 48.3% | 25.50% |
| Use of carsharing within the next year (%) | 36.50% | 10.00% | 4.80% |

## CONCLUSIONS

This study looks into the different likelihood of Californians towards the adoption of AVs and use of self-driving mobility services. Our study uses data collected with the 2018 California Transportation Survey and applies the K-means method to uncover seven clusters: *Multitaskers/ environmentalists/ impaired drivers*, *Tech mavens/ travelers, Life in transition, Captive car-users, Public/ active transport users, Sub-urban Dwellers*, and *Car enthusiasts*. The first two clusters include early adopters who are familiar with the concept of AVs, are tech savvy, and are interested in using the productively during their commute. At the other end of the spectrum, the cluster comprise of non-adopters who are *Car enthusiasts*. People who are *Life in transition*, *Captive car-users*, *Public/ active transport users*, and *Sub-urban dwellers* show uncertain perceptions towards being AV adopters. Interesting, across all groups we find that the interest in being an early AV owner is positively correlated with the interest in being users of self-driving mobility services, though some nuances in this relationship exist across the members of the various clusters.

The results show that the current travel behaviors significantly relate to the AV propensities of the *Multitaskers/ environmentalists/ impaired drivers, Public/active transport users,* and *Sub-urban dwellers*. Majority of *Multitaskers/ environmentalists/ impaired drivers* who grew up and currently live in urban areas own vehicles, but still frequently use public transport, ridehailing, and participate in active transport. Whereas residence type and housing tenure as well as events in life connect with the AV adoption decisions of people who are *Life in transition.*

The findings from this research are helpful for stakeholders, such as AV manufacturing companies, TNCs, and policy makers: for example, it helps to inform AV manufacturers on how to shape their business strategies and target project development to various groups in the population who might turn into potential AV owners. We also identify several segments of latent demand waiting to be tapped by AV manufacturing companies or TNCs: those segments include *Life in transition, Captive car-users, Public/ active transport users,* and *Sub-urban Dwellers*. The findings also support policy makers and can inform current and future regulatory efforts, for example to support the development of policy frameworks to support higher-occupancy on-demand transportation services for those people who currently depend on cars.

The goal of our research is to provide a better understanding and help build a better ecosystem around AV adoption. This study brings a more in-depth understanding of who are the likely, unlikely, and hesitant AV adopters, and what are the characteristics of these groups. We discuss the factors that associate adopter's decisions as well as ways to unlock latent untapped demand to support AV adoption. Nevertheless, several uncertainties affect the study and call for further research. In the nest stages of the research, we plan to explore relationships between AV adoption and interest in micro-mobility. Future transportation systems and Mobility as a Service (MaaS) solutions will revolutionize the transportation landscape, affecting individuals' preferences and propensities towards the use of certain transportation modes. Thus, the dualism between "owning vs. sharing" an AV proposed in certain parts of this research might be short-lived. More complex mobility packages and subscription models are likely to reshape future vehicle ownership and transportation patterns. These are likely to include eventual blending of forms of private vs. shared mobility, as well as they will reshape the nature of on-demand transportation services. Also, travel mode choice behavior is likely to change due to the availability of new AV options as well as purposely-designed vehicles that are optimized for either individual or shared use, whose popularity and attractiveness among travelers should be studied. Another direction is to explore AV adoption behaviors of minority groups.

## ACKNOWLEDGMENTS

The survey design, its administration, and data processing were funded by the National Center for Sustainable Transportation (NCST), which receives funding from the USDOT and Caltrans through the University Transportation Centers program. The authors would like to thank Caltrans, the NCST and USDOT, for their support of university-based research in transportation. Many colleagues contributed to the survey design, data collection, data management and analyses including Grant Mason, Patricia Mokhtarian, Susan Handy, Farzad Alemi, Jai Malik, Yongsung Lee, Ali Etezady and Niloufar Yousefi.

## AUTHOR CONTRIBUTIONS

The authors confirm contribution to the paper as follows: study conception and design: Le; data collection: Circella; analysis: Le; interpretation of results and draft manuscript preparation: Le and Circella. All authors reviewed the results and approved the final version of the manuscript.